\documentclass[runningheads]{llncs}
\usepackage[T1]{fontenc}
\usepackage{hyperref}
\usepackage{booktabs}
\usepackage{amsfonts} 
\usepackage{amsmath} 
\usepackage{graphicx}
\usepackage{multirow}
\usepackage{float}
\usepackage[table]{xcolor}
\usepackage{cleveref} 

\usepackage{graphicx}
\usepackage{color}

\usepackage{tikz}
\definecolor{NaiveColor}{RGB}{0, 114, 178}       
\definecolor{ADcolor}{RGB}{230, 159, 0}       
\definecolor{SDcolor}{RGB}{0, 158, 115}       
\definecolor{CDcolor}{RGB}{204, 121, 167}    
\definecolor{IoUcolor}{RGB}{240, 228, 66}
\definecolor{WholeImageColor}{RGB}{213, 94, 0}     
\definecolor{TheoreticalColor}{RGB}{128, 128, 128}
\definecolor{CLIPcolor}{RGB}{86, 180, 233}  
\definecolor{lightgreen}{RGB}{200, 255, 200}

\newcommand{\MethodBox}[2]{%
  \raisebox{0.5pt}{\textcolor{#1}{\rule{1.2ex}{1.2ex}}}%
  \hspace{0.4ex}\textbf{\textcolor{black}{#2}}%
}

\newcommand{\GNINE}{\MethodBox{NaiveColor}{\textbf{Static-9}}}
\newcommand{\GAPPENDL}{\MethodBox{ADcolor}{\textbf{Append-Long}}}
\newcommand{\GAPPENDS}{\MethodBox{SDcolor}{\textbf{Append-Short}}}
\newcommand{\GFIVE}{\MethodBox{CDcolor}{\textbf{Static-5}}}
\newcommand{\GWHOLE}{\MethodBox{WholeImageColor}{\textbf{Whole-Image}}}
\newcommand{\GTHEORETICAL}{\MethodBox{TheoreticalColor}{\textbf{Theoretical}}}

\begin{document}
\title{Experimental Evaluation of Static Image Sub-Region-Based Search Models Using CLIP}
\titlerunning{Static Image sub-region Search Models}
%
\author{Bastian Jäckl\inst{1}\orcidID{0009-0004-3341-1524} \and
Vojt\v{e}ch Kloda\inst{2}\orcidID{0009-0003-2733-7690} \and Daniel A. Keim\inst{1}\orcidID{0000-0001-7966-9740} \and
Jakub Loko\v{c}\inst{2}\orcidID{0000-0002-3558-4144}}

\institute{University of Konstanz, Konstanz, Germany \and
Charles University, Prague, Czechia}
\authorrunning{Jäckl et al.}
\maketitle              
\begin{abstract}
Advances in multimodal text-image models have enabled effective text-based querying in extensive image collections. While these models show convincing performance for everyday life scenes, querying in highly homogeneous, specialized domains remains challenging. The primary problem is that users can often provide only vague textual descriptions as they lack expert knowledge to discriminate between homogenous entities. This work investigates whether adding location-based prompts to complement these vague text queries can enhance retrieval performance. Specifically, we collected a dataset of 741 human annotations, each containing short and long textual descriptions and bounding boxes indicating regions of interest in challenging underwater scenes. Using these annotations, we evaluate the performance of CLIP when queried on various static sub-regions of images compared to the full image. Our results show that both a simple 3-by-3 partitioning and a 5-grid overlap significantly improve retrieval effectiveness and remain robust to perturbations of the annotation box. 
\keywords{Sub-region search \and Marine images \and CLIP.}\end{abstract}

\section{Introduction}
Multimodal deep learning networks are a cornerstone of modern retrieval systems \cite{clip}. They allow users to query for complex information by writing simple text queries. This approach showed great success in image and video datasets containing everyday life scenes. As a result of this convincing effectiveness, CLIP-based architectures have also become the foundation of interactive video retrieval systems used in academic competitions, such as the Video Browser Showdown (VBS) \cite{vbsEval,vbs2024_eval_performance}. However, these competitions have also revealed severe limitations of CLIP-powered backbones: in highly specific and homogeneous datasets, the effectiveness tends to drop compared to much larger collections like V3C \cite{RossettoSAB19}.

Our work focuses on improving retrieval \emph{for highly homogeneous domains} while still relying on text queries. We chose the Marine Video Kit (MVK) \cite{MVK}, a video dataset that contains more than 28 hours of underwater footage. Specifically, we use a set of preprocessed keyframes that was also used by teams in the VBS \cite{stroh2025prak}. MVK is characterized by highly similar, predominantly blue color tones and features marine life. Some videos contain only unremarkable seafloor, algae, or rocks, making effective text-based querying particularly difficult. In this work, we investigate whether complementing text queries with information indicating \textbf{where} the query should match can enhance retrieval performance. This approach is especially beneficial in scenarios where users can only formulate vague text queries for specific image subregions. 

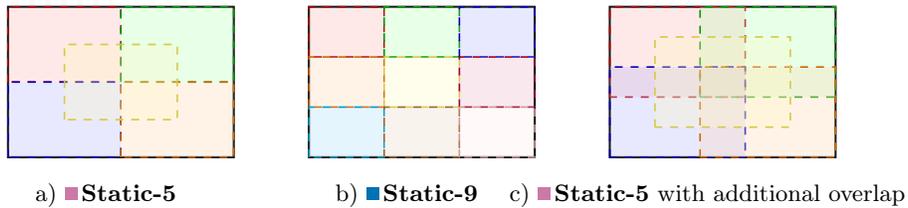
\begin{figure}
\centering
\begin{tikzpicture}[scale=1]

\draw[thick] (0,0) rectangle (3,2);

\def\rw{1.5}
\def\rh{1.0}

\draw[fill=red!30, fill opacity=0.3, draw=red!70!black, semithick, dashed] 
  (0, 1) rectangle (1.5, 2);

\draw[fill=green!30, fill opacity=0.3, draw=green!60!black, semithick, dashed] 
  (1.5, 1) rectangle (3, 2);

\draw[fill=blue!30, fill opacity=0.3, draw=blue!70!black, semithick, dashed] 
  (0, 0) rectangle (1.5, 1);

\draw[fill=orange!30, fill opacity=0.3, draw=orange!80!black, semithick, dashed] 
  (1.5, 0) rectangle (3, 1);

\draw[fill=yellow!30, fill opacity=0.3, draw=yellow!80!black, semithick, dashed] 
  (0.75, 0.5) rectangle (2.25, 1.5);

\node at (1.3,-0.5) {a) \GFIVE};

\draw[thick] (4,0) rectangle (7,2);

\def\colors{{"red!30","green!30","blue!30","orange!30","yellow!30","purple!30","cyan!30","brown!30","pink!30"}}
\def\drawcolors{{"red!70!black","green!60!black","blue!70!black","orange!80!black","yellow!80!black","purple!80!black","cyan!80!black","brown!80!black","pink!80!black"}}

\foreach \i in {0,...,8} {
  \pgfmathsetmacro{\x}{int(Mod(\i,3))}
  \pgfmathsetmacro{\y}{int(2 - floor(\i/3))}

  \pgfmathparse{\colors[\i]} \let\thiscolor\pgfmathresult
  \pgfmathparse{\drawcolors[\i]} \let\thisdraw\pgfmathresult

  \draw[fill=\thiscolor, fill opacity=0.3, draw=\thisdraw, semithick, dashed]
    ({4 + \x}, {2/3 * \y}) rectangle ({4 + \x + 1}, {2/3 * \y + 2/3});
}

\node at (5.3,-0.5) {b) \GNINE};

\begin{scope}[xshift=8cm]
\draw[thick] (0,0) rectangle (3,2);

\def\rw{1.8}
\def\rh{1.2}

\draw[fill=red!30, fill opacity=0.3, draw=red!70!black, semithick, dashed] 
  (0.9 - \rw/2, 1.4 - \rh/2) rectangle 
  (0.9 + \rw/2, 1.4 + \rh/2);

\draw[fill=green!30, fill opacity=0.3, draw=green!60!black, semithick, dashed] 
  (2.1 - \rw/2, 1.4 - \rh/2) rectangle 
  (2.1 + \rw/2, 1.4 + \rh/2);

\draw[fill=blue!30, fill opacity=0.3, draw=blue!70!black, semithick, dashed] 
  (0.9 - \rw/2, 0.6 - \rh/2) rectangle 
  (0.9 + \rw/2, 0.6 + \rh/2);

\draw[fill=orange!30, fill opacity=0.3, draw=orange!80!black, semithick, dashed] 
  (2.1 - \rw/2, 0.6 - \rh/2) rectangle 
  (2.1 + \rw/2, 0.6 + \rh/2);

\draw[fill=yellow!30, fill opacity=0.3, draw=yellow!80!black, semithick, dashed] 
  (1.5 - \rw/2, 1.0 - \rh/2) rectangle 
  (1.5 + \rw/2, 1.0 + \rh/2);

\node at (1.3,-0.5) {c)  \GFIVE\ with additional overlap};
\end{scope}

\end{tikzpicture}
\caption{Three static partitioning options: a) 2x2 grid with an additional center rectangle, b) simple 3x3 grid, and c) 2x2 grid with an additional center rectangle and overlapping regions.}
\label{fig:all-three}
\end{figure}

In general, the efficacy of including position information is twofold: first, it acts as a filter by excluding candidates that contain the queried content but in a different location; second, it reduces the negative influence of image regions outside the bounding box that the user may not describe accurately in the text query. This work incorporates position information by partitioning image candidates into static subregions (see \autoref{fig:all-three}). We then match the text queries only against embedded subregions that match the specified position constraints of the query. Thereby, static grids bear the advantage that they are simple to implement and query, for example, by supplementing the text description with a suffix such as "top right." that constraints the region. To test the retrieval effectiveness of position-enhanced text queries, we collected 741 annotations from 12 annotators for our experiments, including (short and long) descriptions and bounding boxes of the regions that the text queries describe. Our key findings are:

\begin{itemize}
    \item We show that static spatial partitioning of candidate images significantly improves retrieval performance in homogeneous domains like underwater footage, as it enables more precise alignment between text queries and localized visual content.
    \item We find that enriching text queries with positional constraints (e.g., “top left”) is insufficient for effective retrieval
    \item We demonstrate that overlapping grid cells enhance robustness to spatial noise, such as misaligned or imprecise position constraints, making retrieval more reliable under real-world conditions.
\end{itemize}

\section{Related Work} 
Sub-region search is a well-established technique for image and video retrieval. The earliest approaches date back more than 25 years \cite{visualSeek}, using low-level features. Those low-level features are predominantly texture and color, which are readily used for dynamic partitioning and querying \cite{visualSeek,signatureVideoBrowser}. However, some researchers also rely on static approaches for sub-region search, sometimes in combination with features extracted from neural networks \cite{locality_clothes,lineIT}. With the advent of object detectors and segmentation models, researchers used those to partition grids in semantically meaningful regions \cite{spatialSemanticSearch,regionBasedRetrieval,spatialSemanticSearch,conceptMap}. However, these approaches do not allow for natural language queries but rely on fixed-object categories or querying by example.

Powerful multimodal networks, such as CLIP \cite{clip}, enabled researchers to build retrieval systems with natural language queries. While they allow the formulation of position constraints in the textual query, they are not well-optimized for these constraints \cite{badLocalization}. Recent efforts have thus explored more sophisticated pipelines built on top of vision foundation models (VFMs). Shlapentokh-Rothman et al.~\cite{regionBasedRepresentations} revisit region-based image representations by integrating segmentation masks generated by SAM~\cite{kirillov2023segany} with features extracted from DINOv2~\cite{dinov2}. They demonstrate that simply pooling features over segmented regions can yield competitive performance on semantic segmentation and object-centric retrieval tasks while supporting efficient and flexible image search. Finally, Search Anything~\cite{searchAnything}, a zero-shot image region retrieval method, leverages FastSAM~\cite{fastSam} for segmentation and combines it with CLIP to generate binary hash codes for fine-grained, region-level similarity search without requiring labeled data. Search Anything requires an image-based query.

Due to the large success of sub-region-based search systems, researchers also incorporated the functionality in academic competitions such as the VBS \cite{LokocBSD21,LokocMVS21}. In the latest iteration of the VBS, Stroh et al. \cite{stroh2025prak} found success by incorporating static grid partitioning. Users of the system can query for one of five regions by appending a position suffix to the text query. They also offer a preliminary evaluation of various static partitioning models - we offer a more thorough evaluation in the subsequent chapters.

\section{Data Collection}\label{sec:data}
For our experiments, we collected a total of 741 annotations. Each annotation contains a short textual description, a long description, and a bounding box for a region of interest. This section describes the collection procedure and offers detailed statistics of the dataset.

\subsection{Collection Procedure}
\begin{figure}
    \centering
    \includegraphics[width=\linewidth]{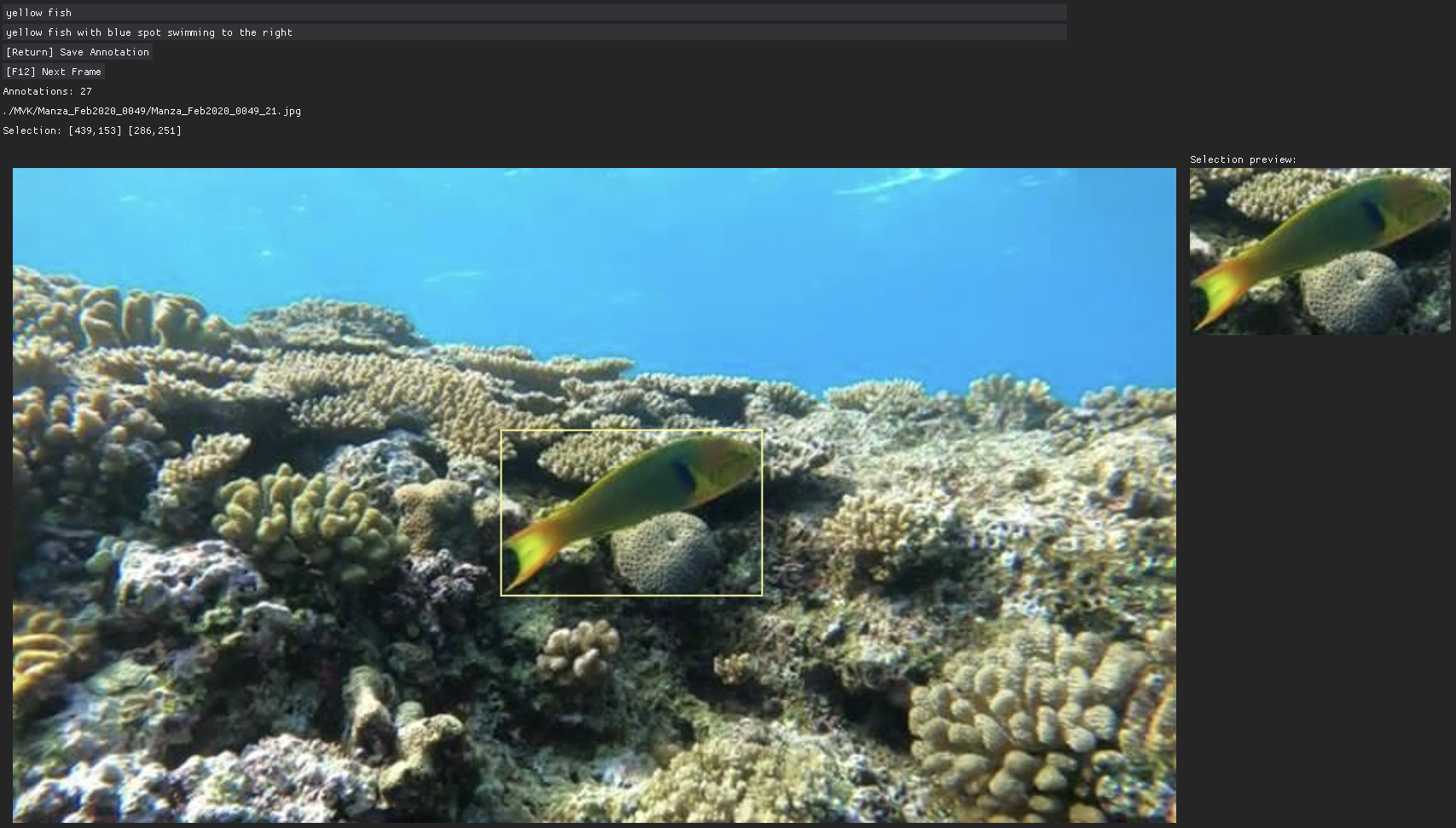}
    \caption{The annotation tool used for our work. Annotators could draw bounding boxes on a canvas with images, and were asked to write a short and long textual descriptions.}
    \label{fig:annotation_tool}
\end{figure}

We collected 741 annotations from 12 annotators (ten male, two female), aged between 23 and 43 years ($\mu= 26.5$ , $\sigma=6.2$). None of the annotators are domain experts, but 6 out of 12 already worked scientifically with the MVK dataset. To ease the collection process, we implemented a tool with a simple user interface, exemplary displayed in \autoref{fig:annotation_tool}. The tool prompts annotators with (previously extracted) keyframes from the MVK dataset. The set of keyframes was used by teams in VBS \cite{stroh2025prak}. Subsequently, annotators are asked to provide short and long textual descriptions of interesting regions in the keyframe and draw a bounding box around it. Annotators were allowed to annotate multiple regions per image. We had no further constraints except limiting the maximum size of bounding boxes to 30\% of the original image. All annotators executed two runs: In the first run, annotators could skip images where they did not observe interesting regions. In the second run, annotators did not have that option and were consequently forced to provide at least one annotation per image. We refer to the subsets as "skippable" and "non-skippable" for the remainder of this work. We required that the annotators provide at least 20 annotations. Remarkably, in our experimental setup, annotators draw the bounding box directly into the keyframes (see \autoref{fig:annotation_tool}), allowing for pixel-accurate rectangles. We will discuss the impact of these "optimal" bounding boxes in the next subsection.

\subsection{Optimal Bounding Boxes}
 In practice, users usually do not know the exact position of the region they are querying for. Even in tasks where the target frame is given, such as visual known-item search tasks, it is impossible to draw pixel-accurate regions on a blank canvas. We nevertheless consider the collected "optimal" bounding boxes essential as they provide an upper-bound limit for retrieval accuracy with static partitionings. Furthermore, we merely use the rectangles as \emph{estimate} to select predefined subregions to query in. We evaluate in \autoref{sec:robustness} that static grids are robust to small to medium-sized perturbations of position and size of these bounding boxes.

\subsection{Annotation Statistics}
We provide an overview of the collected annotations in \autoref{tab:query-annotations}. Description lengths, rectangle sizes, and position of rectangles are comparable for skippable and non-skippable annotations.

\begin{table}[ht]
\centering
\begin{tabular}{l@{\hskip 10pt}r@{\hskip 10pt}c@{\hskip 10pt}c@{\hskip 10pt}c}
\toprule
\textbf{} & \textbf{Count} & \textbf{Short Desc.} & \textbf{Long Desc.} & \textbf{Rect. Size} \\
\midrule
Skippable      & 486 &      $12.91\pm 6.96$       &    $46.15 \pm 19.35$                &  $19.0 \pm 19.3$                                                             \\
Non-Skippable  & 255 &      $12.62 \pm 6.87$       &    $41.77 \pm 18.92 $                &    $16.4  \pm 18.3$                                                                \\
\midrule
Total           &   741          &     $12.81\pm 6.93$               &       $44.65 \pm 19.32$              &   $18.1 \pm 19.1  $                                          \\
\bottomrule
\end{tabular}
\caption{Overview of skippable and non-skippable query annotations. "Short. Desc." counts the average number of characters in the short text queries, "Long. Desc." analogously counts the characters for long text queries. "Rect. Size." measures the average area (percentage) of the image that is occupied by the annotation bounding boxes.}
\label{tab:query-annotations}
\end{table}

\begin{figure}
    \centering
    \includegraphics[width=1\linewidth]{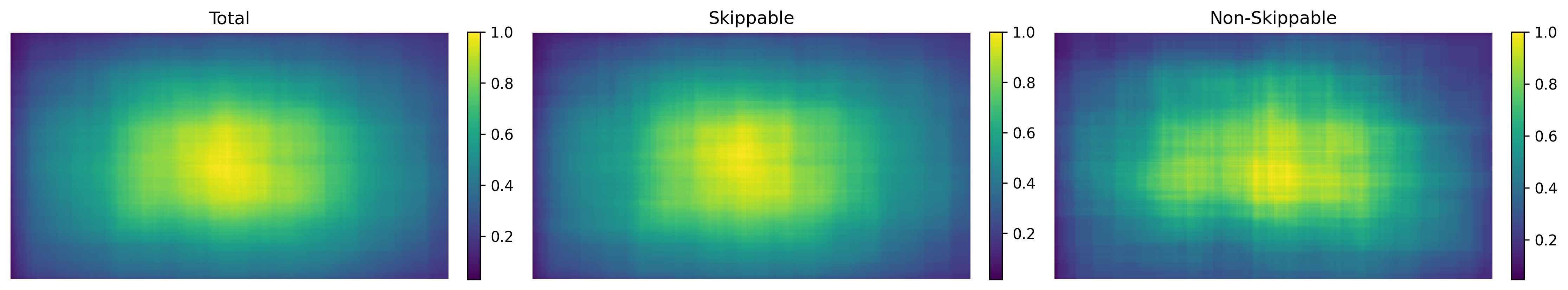}
    \caption{Heatmaps showing the spatial distribution of annotated rectangles across the dataset. From left to right: all rectangles, rectangles from skippable annotations, and rectangles from non-skippable annotations. Brighter regions indicate areas with higher annotation density.}
    \label{fig:heatmap_rectangles}
\end{figure}

We show the annotation rectangles distribution in \autoref{fig:heatmap_rectangles}. Most annotations are placed in the middle of the image and fade out towards the corners. 

\section{Methods}\label{sec:method}
We use the collected annotations to investigate the retrieval effectiveness of partitioning grids. In this section, we first establish two baselines, namely the whole-image baseline and a theoretical upper-bound baseline, that is unfeasible in practice. Finally, we formally introduce the matching and ranking of static partitioning search models.

\subsection{CLIP-based Image Retrieval} \label{whole-image-computation}
Recent interactive video retrieval tools predominantly allow users to formulate textual queries to retrieve a set of keyframes. Keyframes are extracted in a preprocessing pipeline alongside their embeddings for efficiency. Specifically, for every keyframe $k_i$ in a set of keyframes $K$, features $f_{k_i} \in \mathbb{R}^d$ are computed using CLIP's image embedder. Thus, $K$ is represented by the feature matrix $F_K \in \mathbb{R}^{n \times d}$ for $n$ keyframes. Then, during query time, a text query $t$ is embedded with CLIP's text embedder, resulting in the feature $f_t  \in \mathbb{R}^d$. Assuming $f_t$ and $F_K$ are normalized, the ranking scores for all keyframes can be computed via matrix multiplication:

\begin{equation}
    \mathbf{s} = F_K f_t^\top
\end{equation}

Here, $\mathbf{s} \in \mathbb{R}^n$ contains the cosine similarity scores between the query and each keyframe. Keyframes are then ranked in descending order of their respective scores, enabling efficient retrieval based on semantic similarity. We denote this standard approach as \GWHOLE. Because annotators were instructed to describe only specific regions of interest—excluding surrounding context—the whole-image baseline naturally includes less relevant information than the partition-based approaches. To enable a fair comparison, we enrich the text queries with location indicators. Concretely, we determine which of the five bounding boxes in the 5-image grid has the highest IoU with the annotated region and append a corresponding positional phrase. For example, if the top-left box has the highest IoU, we append "in the upper left" for \GAPPENDS\ and "in the upper left part of the image" for \GAPPENDL.

\subsection{Theoretical Upper Bound}
In addition to the commonly used full-frame baseline, we introduce \GTHEORETICAL\ as an upper-bound baseline. \GTHEORETICAL, while computationally infeasible in practice, estimates the potential of region-based retrieval with perfect bounding boxes. Instead of relying on precomputed image features, this method performs retrieval at query time using localized crops derived from annotations.

Given a specific annotation bounding box \( b_i = (\text{left}_i, \text{top}_i, \text{width}_i, \text{height}_i) \), we crop \emph{all} keyframes \( k_j \in K \) using the same coordinates \( b_i \), resulting in a modified dataset \( \mathcal{D}_{\text{crop}}^{(i)} \) consisting of tightly focused regions. Each cropped region is then passed through CLIP's image encoder to obtain features \( f^{\text{crop}}_{k_j, i} \in \mathbb{R}^d \). As before, the text query is embedded as \( f_t \in \mathbb{R}^d \).

Assuming all embeddings are normalized, we compute the cosine similarity scores via matrix multiplication:

\begin{equation}
    \mathbf{s}^{(i)} = F_{\text{crop}}^{(i)} f_t^\top
\end{equation}

where \( F_{\text{crop}}^{(i)} \in \mathbb{R}^{n \times d} \) contains the features \( f^{\text{crop}}_{k_j, i} \) for all keyframes \( k_j \). This process is repeated for each annotation \( i \), effectively simulating an oracle that always selects the ideal centered region for comparison. While this setup is infeasible for real-time systems due to the need for per-query cropping and embedding computation, it provides an upper bound on the retrieval performance achievable with perfect region localization.

\subsection{Static Sub-region based Image Retrieval}


Static partitioning-based image retrieval extends \GWHOLE\ by evaluating queries on sub-regions. This is a two-step process:

Let each image \( I \) be partitioned into a fixed set of sub-regions \( R_I = \{ r_1, \dots, r_m \} \) defined by a static spatial grid. Let \( B_I \subseteq R_I \) denote the subset of regions that intersect with the annotated bounding box of interest in image \( I \). The intersection condition is defined as:
\[
r_j \in B_I \iff \mathrm{IoU}(r_j, b) > 0
\]
for a given annotation box \( b \) in image \( I \), where IoU denotes the intersection-over-union. (Alternatively, a stricter selection could be used, such as selecting only \( \arg\max_{r_j \in R_I} \mathrm{IoU}(r_j, b) \).)

Each sub-region \( r_j \in R_I \) is associated with a pre-computed CLIP feature vector \( f_{r_j} \in \mathbb{R}^d \). These are stacked into a matrix \( F_R \in \mathbb{R}^{m \times d} \) for each image. Given a normalized query embedding \( f_t \in \mathbb{R}^d \), the similarity scores for all subregions in image \( I \) are computed as:
\[
\mathbf{s}_I = F_R f_t^\top \in \mathbb{R}^m
\]

From this, we define the \emph{image-level score} as the highest similarity among its relevant sub-regions:
\[
s_I = \max_{r_j \in B_I} \mathbf{s}_I[j] \in \mathbb{R}
\]

Similar to the whole-image baseline, the similarity scores yield a ranking over images.

\subsubsection{Partitioning Variants} We consider two static partitioning strategies and variations of those with overlapping rectangles (see \autoref{fig:all-three}). 

For \GNINE\ we define a uniform \(3 \times 3\) grid, where each sub-region \(r_{i,j}\) covers a rectangular area of width \( \frac{1}{3} \) and height \( \frac{1}{3} \) of the original image domain. Formally, the image is partitioned into 9 equal cells (see \autoref{fig:all-three}, b). The grid, specified in normalized coordinates \((x_1, y_1, x_2, y_2)\), are:
\[
R_{9\text{-grid}} = \left\{ \left( \frac{i-1}{3}, \frac{j-1}{3}, \frac{i}{3}, \frac{j}{3} \right) \;\middle|\; i, j \in \{1,2,3\} \right\}
\]

Second, we define the \GFIVE\ layout consisting of four corner regions and one central region (see \autoref{fig:all-three}, a). Each corner region covers one-quarter of the image, and the central region is defined to cover the middle half of the image in both width and height:
\[
r_{\text{top-left}} = [0, 0, \frac{1}{2}, \frac{1}{2}], \quad
r_{\text{top-right}} = [\frac{1}{2}, 0, 1, \frac{1}{2}]
\]
\[
r_{\text{bottom-left}} = [0, \frac{1}{2}, \frac{1}{2}, 1], \quad
r_{\text{bottom-right}} = [\frac{1}{2}, \frac{1}{2}, 1, 1]
\]
\[
r_{\text{center}} = [\frac{1}{4}, \frac{1}{4}, \frac{3}{4}, \frac{3}{4}]
\]
\[
R_{\text{5-grid}} = \{ r_{\text{top-left}}, r_{\text{top-right}}, r_{\text{bottom-left}}, r_{\text{bottom-right}}, r_{\text{center}} \}
\]

\section{Experiments}
We use the annotation set described in \autoref{sec:data} to test the retrieval effectiveness. Specifically, we measure the rank of the target image in the set of MVK keyframes with the previously introduced grid partitionings. We provide an overview of the results in  \autoref{tab:overview} where we report the recall R and the mean rank (MNR). We subsequently discuss those numbers and enrich them with further insight.

\begin{table}
    \centering
    \caption{Retrieval performance for Skippable and Non-Skippable annotations. We report the recall R@rank in \% and the mean rank (MNR). We highlight the best performing grid partitioning method (excluding the theoretical upper-bound).}
    \resizebox{\textwidth}{!}{
    \begin{tabular}{l ||c|c|c|c|c || c|c|c|c|c}
        \toprule
        \multirow{2}{*}{Model} 
        & \multicolumn{5}{c}{Skippable} 
        & \multicolumn{5}{c}{Non-Skippable} \\
        \cmidrule(lr){2-6} \cmidrule(lr){7-11}
        & R@1$\uparrow$& R@10$\uparrow$ & R@100$\uparrow$ & R@1000$\uparrow$ & MNR$\downarrow$ & R@1$\uparrow$ & R@10$\uparrow$ & R@100$\uparrow$ & R@1000$\uparrow$ & MNR$\downarrow$  \\
        \midrule
        \multicolumn{11}{l}{\textbf{Long Queries}} \\
        \GWHOLE        & 0 & 6 & 21 & 47 & 9320 & 0 & 2 & 11 & 34 & 11064 \\
        \midrule
        \GAPPENDL        & 0 & 6 & 19 & 44 & 10514 & 0 & 1 & 8 & 31 & 11357 \\
        \GAPPENDS        & 0 & 6 & 19 & 45 & 10543 & 0 & 2 & 9 & 31 & 11522 \\
        \midrule
        \GFIVE         & 1 & \cellcolor{lightgreen}9 & \cellcolor{lightgreen}29 & 56 & 7715 & 0 & 2 & \cellcolor{lightgreen}18 & \cellcolor{lightgreen}43 & \cellcolor{lightgreen}7419 \\
        \GNINE          & \cellcolor{lightgreen}2 & \cellcolor{lightgreen}9 & \cellcolor{lightgreen}29 & \cellcolor{lightgreen}57 & \cellcolor{lightgreen}7400 & \cellcolor{lightgreen}2 & \cellcolor{lightgreen}3 & 15 & 41 & 8163 \\
        \midrule
        \GTHEORETICAL & 8 & 19 & 41 & 67 & 5684 & 5 & 11 & 30 & 57 & 5447 \\
        \bottomrule
        \bottomrule
        \multicolumn{11}{l}{\textbf{Short Queries}} \\
        \GWHOLE       & 0 & 3 & 15 & 40 & 8861 & 0 & 1 & 10 & 28 & 13242 \\
        \midrule
        \GAPPENDL        & 0 & 3 & 11 & 31 & 12261 & 0 & 1 & 8 & 22 & 14609 \\
        \GAPPENDS        & 0 & 2 & 11 & 30 & 12738 & 0 & 0 & 7 & 20 & 15528 \\
        \midrule
        \GFIVE         & 0 & 4 & 19 & 51 & \cellcolor{lightgreen}7703 & \cellcolor{lightgreen}1 & \cellcolor{lightgreen}4 & \cellcolor{lightgreen}12 & 31 & \cellcolor{lightgreen}9555 \\
        \GNINE          & 0 & \cellcolor{lightgreen}5 & \cellcolor{lightgreen}21 & \cellcolor{lightgreen}52 & 8780 & 0 & 2 & 9 & \cellcolor{lightgreen}35 & 10619 \\
        \midrule
        \GTHEORETICAL & 5 & 10 & 28 & 56 & 8120 & 2 & 5 & 18 & 41 & 9118 \\
        \bottomrule
    \end{tabular}}
    \label{tab:overview}
\end{table}
\subsection{Appending Textual Localization}
We first note that naively appending position information to the textual queries does not work. The retrieval effectiveness of \GAPPENDL\ and \GAPPENDS\ is worse than \GWHOLE. This effect can not be explained by worse target-query matching, as the similarity scores slightly improve for the appended grids (from 0.159 to 0.166 for long suffixes on average). Thus, the lower retrieval accuracy can only be explained by the fact that the append localization information better matches false candidates. Besides that, we note that appending position information by text is not ideal due to its non-specificity and ambiguity.

\subsection{Static Grid Partitionings}
From \autoref{tab:overview}, we observe that both \GFIVE\ and \GNINE\ significantly improve over \GWHOLE. Notably, R@10 and R@100 increase by nearly 50\% for both skippable and non-skippable annotations. With that, they remain far behind the perfect grid partitioning of \GTHEORETICAL. We thus notice that a possible future work is grids that segment images into coherent regions, e.g., with the help of object detectors. However, even with optimal segmentation, the target is only in 41\% of cases in the first 100 retrieved candidates. This highlights the challenge and limitations of text queries as ranking functions.

The performance difference between \GFIVE\ and \GNINE\ is only moderately substantial. In the following, we analyze the reasons for these improvements over the whole-image baseline.The advantages of static grid partitioning are twofold:

\begin{enumerate}
    \item \textbf{Improved localization:} If a grid cell closely aligns with the annotation bounding box, it better captures the relevant visual content described in the text, avoiding irrelevant surrounding information that could otherwise degrade the embedding quality.

    \item \textbf{Noise filtering:} Static partitioning acts as a spatial filter. In whole-image embeddings, irrelevant but textually similar regions from false candidates can affect similarity scores. With grid partitioning, such regions only contribute if they happen to fall within the same grid cell as the annotation, thus reducing false matches.
\end{enumerate}

To test whether static grid cells better align with annotation bounding boxes, we compare their Intersection over Union (IoU) scores. The mean IoU for skippable annotations is 0.33 for \GFIVE\ and 0.31 for \GNINE, compared to just 0.19 for the whole-image baseline. Comparable trends are observed for non-skippable annotations. While these IoU differences are non-negligible, they do not fully account for the substantial performance gains.

We find no strong direct correlation between IoU and retrieval rank improvements (correlation coefficient: $-0.13$ for both grid variants). However, finer-grained embeddings do partially explain the improved ranks, which we will investigate in the following. We start by computing CLIP embeddings between the text query and 1) the whole-image embedding, and 2) the best-matching grid cell embedding (from the partitioned variant).

Formally, let $\mathbf{s}^{\text{whole}}$ and $\mathbf{s}^{\text{part}}$ denote the vectors of CLIP similarities between the text queries and the image embeddings of targets for the whole image and the best-matching partitioned grid cell, respectively. Likewise, let $\mathbf{r}^{\text{whole}}$ and $\mathbf{r}^{\text{part}}$ denote the corresponding vectors of retrieval ranks of the targets. Each vector is of length $N$, the number of annotated queries. While the mean values show seemingly small differences ($\mu (\mathbf{s}^{\text{whole}}) = 0.168$ and $\mu (\mathbf{s}^{\text{part}}) = 0.159$), we find statistical differences using a Wilcoxon signed rank test, with $\mathbf{s}^{\text{part}}$ being significantly higher (N = 486, p < 0.001) - we previously applied a Shapiro–Wilk test to verify that the similarities are not normally distributed.

Subsequently, we compute the Pearson correlation between the element-wise differences $\Delta \mathbf{s} = \mathbf{s}^{\text{whole}} - \mathbf{s}^{\text{part}}$ and $\Delta \mathbf{r} = \mathbf{r}^{\text{whole}} - \mathbf{r}^{\text{part}}$. The correlation between $\Delta \mathbf{s}$ and $\Delta \mathbf{r}$ amounts to $-0.69$. The strong correlation between differences in CLIP similarities and rank, combined with the generally higher CLIP similarities of partition approaches, suggest that improvements in retrieval performance are at least partially driven by more precise text-image alignment.

Finally, we conduct an additional experiment to determine whether the benefits of static partitioning stem solely from filtering out regions of candidates, or from improved embedding quality. To isolate the effect of finer-grained embeddings of targets, we measure the ranks with a slightly modified database. All candidates utilize the whole-image embedding, while only the target uses the finer-grained embedding. To ensure fairness, we only use the best-matching target embedding (based on IoU). We observe substantial performance gains over the baseline. For instance, using \GNINE\ partitionings for the target embedding and evaluating on skippable annotations with long text queries, we achieve R@1 = 0, R@10 = 15, R@100 = 31, and R@1000 = 56. The ranks even keep improving when using the embeddings of the annotation box directly (R@1=0, R@10 = 22, R@100=42, R@1000=66). These results provide strong evidence that static grid partitioning enhances retrieval not just through spatial filtering but also by enabling more accurate text-image alignment.

\subsection{Robustness and Box Enlargements}\label{sec:robustness}
The previously presented experiments represent upper-bounds using perfectly aligned annotation boxes. We now evaluate how robust the methods are to perturbations commonly occurring in real-world scenarios. To this end, we apply two types of perturbations: (1) positional shifts and (2) changes in box size.

For positional shifts, we independently sample horizontal and vertical offsets, \( d_{\text{shiftX}} \) and \( d_{\text{shiftY}} \), from a normal distribution:
\[
d_{\text{shiftX}}, d_{\text{shiftY}} \sim \mathcal{N}(0, \sigma_s),
\]
where \( \sigma_s \) is the standard deviation. These values are added to the original box coordinates to simulate misalignment.

For size perturbations, we alter the width and height by scaling them with factors \( d_{\text{areaX}} \) and \( d_{\text{areaY}} \), drawn from:
\[
d_{\text{areaX}}, d_{\text{areaY}} \sim \mathcal{N}(1, \sigma_a),
\]
where \( \sigma_a \) is the standard deviation. These scale factors adjust the box dimensions while keeping the centroid fixed. After perturbation, boxes are shifted if necessary to ensure they remain within the image frame.

We conduct experiments using specific combinations of the spatial shift standard deviation \(\sigma_s\) and the area scaling standard deviation \(\sigma_a\) to systematically analyze robustness. The selected combinations are: 
\[
(\sigma_s, \sigma_a) \in \{ (0, 0),\ (0.1, 10),\ (0.25, 25),\ (0.5, 50) \}.
\]
The pair \((0, 0)\) represents the unperturbed reference case with perfectly aligned boxes. The other configurations introduce increasing levels of perturbation in both position and size, allowing us to assess how performance degrades under more challenging conditions. We use the same set of perturbations across all grid variants to ensure fairness and further use five seeds per annotation to shrink the bias of random effects.

To further assess robustness, we also evaluate whether enlarging the grid partitioning cells can compensate for shifted or enlarged annotations. Therefore, we define a fixed increase (relative to the whole frame) applied to each grid cell. Border grid cells are always enlarged inwards, while interior cells are uniformly spread in all directions (for half the amount), so each grid cell has the same area after the enlargement. An example is shown in \autoref{fig:all-three}, c). The enlargement increases the likelihood that relevant content remains within the embedding region, even when annotations are perturbed. We test multiple enlargement levels of the original cell size and measure how retrieval performance responds. 

\begin{figure}
    \centering
    \includegraphics[width=1\linewidth]{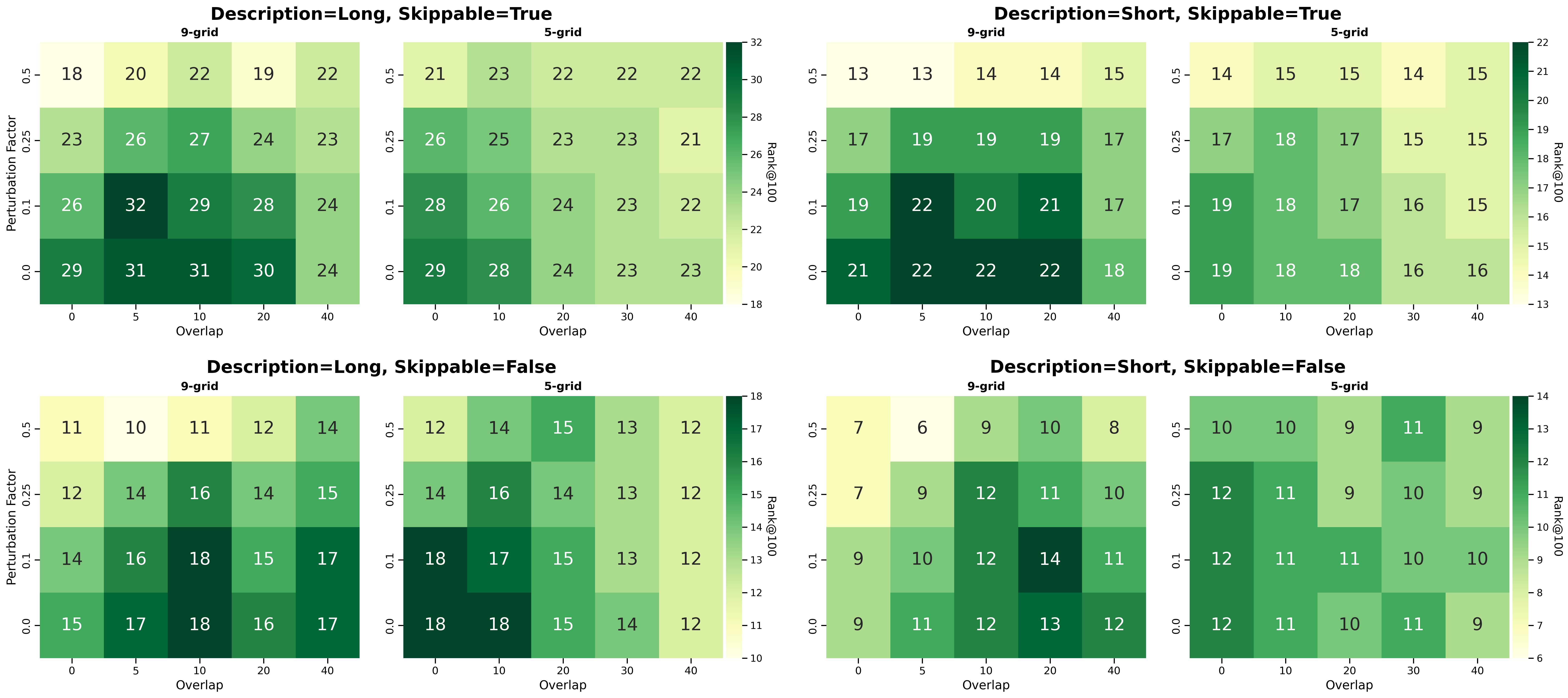}
    \caption{Impact of perturbation factor and spatial overlap of grids on recall R@100 performance across different grid configurations. The reported "Overlap" is the enlargement \% relative to the whole-grid.}
    \label{fig:heatmap}
\end{figure}

We present the results of our robustness analysis in \autoref{fig:heatmap}. Interestingly, for the unperturbed case (0,0), spatial overlaps lead to noticeable improvements in Rank@100, especially for the \GNINE\ setting. This effect can be attributed to the smaller grid cells of the \GNINE\ layout: annotation boxes tend to align only partially with such small partitions. Introducing a slight spatial overlap helps better capture the visual content associated with the annotation, thereby improving retrieval. In contrast, the \GFIVE\ setting, which has larger cells, shows optimal performance with little or no overlap—suggesting that excessive overlap may introduce unnecessary noise in coarser partitionings. Across configurations, performance generally peaks at moderate overlap levels (10–20\%) and tends to decline beyond that.

All grid-based approaches exhibit strong robustness to minor perturbations ($\sigma_s = 0.1$, $\sigma_a = 10$), with only slight degradation in performance. However, performance declines at higher perturbation levels ($\sigma_s = 0.25$, $\sigma_a = 25$), where annotations are substantially shifted or resized. In these scenarios, the use of overlapping grid cells proves especially advantageous. Notably, the \GNINE\ variant with moderate overlaps maintains higher performance than other settings, indicating that finer grids benefit more from overlap when faced with spatial uncertainty. This suggests that overlapping grid partitionings not only improve alignment in ideal cases but also provide resilience under real-world conditions.

\section{Conclusions}
This paper presents a study focusing on known-item search in a highly homogeneous marine dataset. The investigated search task is restricted in two ways. First, participating users are not experts in the marine domain, and thus, the collected text queries are more abstract. Second, users describe only a sub-region of the searched image, including the region rectangle specification of the described sub-region. In order to improve search effectiveness, two types of static partitioning approaches are tested for provided localized text queries. Experiments reveal that static partitioning substantially improves ranking performance compared to the default search mode in the whole image. Surprisingly, using the location information just in the text query prompt turned out to be less effective than searching in the whole image without the location information. We further introduced a theoretical oracle baseline that uses annotation boxes to crop the candidates. While this baseline showed the potential of sub-image search with perfectly aligned grid cells, it also revealed an important limitation: Although it used perfect grid cells, in less than 50\% of the cases targets were in the first 1000 retrieved candidates.

Additionally, we tested the robustness of static grids to imprecise annotation boxes. With simulated perturbations of query region, both static partitioning approaches benefited from small partition extensions leading to overlaps. However, with the introduced misalignment of the query region and semantic description, the ranking performance may drop up to 65\%. Therefore, it is essential to remember the sufficiently precise position and size of the described query region to benefit from the static partitioning approach. This assumption leads to one serious limitation of the presented study. The study was performed with benchmark data collected from users who observed the annotated image during the annotation process. All provided regions and text descriptions benefit from a "perfect memory" assumption, which is indeed not realistic. Therefore, all findings represent an upper-bound performance estimate, and further studies are necessary. As future work, we plan to collect localized user queries in more realistic settings, for example, in VBS-like benchmark evaluations where users memorize played scenes and then focus on query specification in a canvas component.

%
%
\bibliographystyle{splncs04}
\bibliography{paper}

\end{document}